\documentstyle[aps,preprint]{revtex}

\def\lab#1{\label{#1} }

\def\cit#1{\cite{#1}}
\def\beq{\begin{equation}}
\def\eeq{\end{equation}}
\def\bea{\begin{eqnarray}}
\def\eea{\end{eqnarray}}
\def\e{{\mathrm e}}
\def\tr{{\mathrm Tr}}
\def\0{\over}
\def\om{\omega}
\def\la{\lambda}
\def\de{\delta}
\def\li{\left\langle}
\def\re{\right\rangle}
\def\half{ {}^1 \hspace*{-0.2em} /_2 }
\def\ad{a^{\dagger}{} }
\def\ii{{\mathrm i}}
\def\sn{{\mathrm sn}}
\def\un{1 \hspace*{-1ex} 1 }
\def\qd{\quad }
\def\lb{\left[ }
\def\rb{\right] }
\def\CS{{\cal S }}

\begin{document}
\draft
\preprint{{\small E}N{\large S}{\Large L}{\large A}P{\small P}-A-546/95,
hep-ph/9509302}

\title{The dynamics of a thermal non-equilibrium anharmonic oscillator}

\author{Herbert Nachbagauer
\footnote[3]{e-mail: herby @ lapphp1.in2p3.fr}}

\address{Laboratoire de Physique Th\'eorique
ENSLAPP  \footnote{URA 14-36 du CNRS, associ\'ee \`a l'E.N.S. de Lyon,
et au L.A.P.P. (IN2P3-CNRS) d'Annecy-le-Vieux}
Chemin de Bellevue, BP 110, \\ F - 74941 Annecy-le-Vieux Cedex,
France}

\date{\today}

\maketitle

\begin{abstract}
We propose an non-standard method to calculate non-equilibrium physical
observables. Considering the generic example of an anharmonic quantum
oscillator, we take advantage of the fact that the commutator algebra of
second order polynomials in creation/annihilation operators closes.
 We  solve the von~Neumann
equation for the density-operator exactly in the mean field
approximation and study the time evolution of the particle number and
the expectation value $\left< X^2\right>  $.

\end{abstract}


\section*{Introduction }

The consistent calculation of equilibrium finite-temperature physical
observables requires a resummation of the leading
temperature contributions along the lines of the Braaten-Pisarski
\cit{Braa} resummation program established already some years ago.
The application of those ideas to scalar theories, QED, QCD and the
electroweak model has been extremely successful so far, although the
original resummation may have to be modified slightly near the
light-cone for models involving massless particles.
On the other hand, many interesting realistic physical scenarios like,
e.g., phase  transitions \cit{phase} and particle production in a plasma
are clearly out-of-equilibrium or only nearly-equilibrium systems in
which case the field theoretical framework lacks a systematic basis
and the straight-forward application of basically linear response methods
gives rise to potential inconsistencies, such as pinch singularities
\cit{pinch}. In this note we consider the toy-model of a
quantum-mechanically anharmonic oscillator which can be solved exactly.

\section*{The thermal anharmonic oscillator}

The model under consideration is characterized by the  Hamiltonian
\beq  H   = H_0 + H_I, \quad  H_0  = {1 \0 2} P^2 + {\om^2 \0 2 } X^2
, \quad  H_I  =  {\la \om^3 \0 2 }  X^4 , \eeq
the quartic interaction-therm  mimicking a
self-interaction term as in the field-theoretical analog.
Since there quadratic Hamiltonians are the only ones
which can be integrated out exactly, we will also here  focus on the
so-called mean-field approximation. The basic idea
is to expand the Hamiltonian operator around an expectation
value that at finite temperature $T$ (and for mixed states in general)
is given by the trace of the product of the corresponding
operator with the normalized density-operator of the state.
For the Hamiltonian above being parity
even, the natural expansion parameter is
$\de X^2 = X^2 -  \li X^2 \re  .$ Dropping the
$( \de X^2 )^2 $ term, the quadratic contribution of the
interaction Hamiltonian,
$H_I^{quadr.} = \half \la \om^3 \li  X^2 \re
( 2 X^2 - \li  X^2 \re),$ has a form
analogous to $H_0$ with time-dependent frequency.
In terms of creation and annihilation operators, $a=\sqrt{2 \om} X +
\ii \sqrt{2/\om} P$ and $ \ad $,
the effective free Hamiltonian $H^{quadr.}$
under study is thus composed of
\beq   H_I^{quadr.}  = {\la \om \0 2 } f(t)
 (\ad^2  + 2 \ad a + a^2 + \un -  f(t) )
\qd \mbox{and} \qd   H_0  = \om  ( \ad a  + {1 \0 2 } )
\lab{quad} \eeq
where $ f(t)/\om = \li X^2 \re = Z^{-1}\, \tr ( \rho  X^2 )$ has to be
determined  self-consistently.
Note that diagonalizing the quadratic Hamiltonian would render the
 vacuum definition $a \left|0\right>=0  $ time dependent.

The density operator for the system in thermal  equilibrium
\beq \rho_{eq} =Z_{eq}^{-1} \exp (-H/T) , \qd Z_{eq}  =
\tr ( \rho_{eq} ) ,\eeq
which minimizes the equilibrium entropy
functional  $S \sim \tr ( \rho \log \rho ),$
is a particular solution of the  von~Neumann  equation of motion
\beq i {{\mathrm d} \0{\mathrm d} t} \rho = \lb H , \rho \rb .  \lab{vn}
\eeq
Solutions of that equation do not uniquely determine the
density-operator but admit in general additive static solutions that are
functionals of eventual constants of motion. I want to point out that
the equilibrium entropy functional remains
constant in time for every solution of (\ref{vn}). For simple, e.g.
zero-dimensional integrable  or infinite dimensional non-interacting,
systems it is argued that thermodynamic information loss has to be induced
by a second averaging procedure, e.g. coarse graining in phase space
\cit{coarse} or in temperature \cit{zubarev}, or  time smoothing.
For more complicated systems as quantum field-theoretical
models or classical  many-body systems the second averaging process is
hidden in asymptotic approximations, and particularly in cutting the
Bogoliubov-Born-Green-Kirkwood-Yvon
hierarchies \cite{hu}, generally resulting in a kinetic equation.
We want to mention that besides those somehow artificial
and pragmatic  methods there exists, based on a couple of evident
assumptions, a systematic approach to clearly distinguish thermodynamic
(the 'relevant') and microdynamic (the 'irrelevant') part of the dynamics
\cit{balescu} which, however, has not yet been applied systematically
within the framework  of quantum
field-theory so far. We will not address these open questions further but
pursue the modest task to find a solution of the von~Neumann equation.

The biquadratic form of (\ref{quad}) in $a,\ad$ suggests the solution $\rho$
\beq \rho = Z^{-1} \exp (B \ad^2 ) \exp ( A \; \ad a + g \un )
\exp ( \bar B a^2 ) ,\qd Z = \tr \rho  , \eeq
where  $A,B$ and $g$
 are time dependent functions to be determined, and
the hermiticity of the density operator $\rho=\rho^{\dag}$
necessitates  $A,g$ real. On account of the basic relation
$\lb a ,\ad \rb = \un $ the set of operators
$\CS = \left\{ \un , \ad a ,a^2 ,\ad^2 \right\} $
forms a closed commutator algebra
\beq \lb \ad^2, a^2 \rb =  - 2 (\un + 2 \ad a ) \qd
\lb \ad^2 ,\ad a \rb = - 2 \ad^2 \qd
\lb a^2,\ad a \rb = 2 a^2 \eeq
which is the essential property that allows the evaluation of the
expectation values in what follows. Using relations such as
$$ \e^{B \ad^2 } a^2 \e^{- B \ad^2 } =
a^2 -2 B (\un +2 \ad a) + 4 B^2 \ad^2 $$
it is elementary algebra to commutate the summands of $H$ to the
r.h.s. of (\ref{vn}) and to compare with their counterparts in the
time derivative. In the resulting system of  ordinary first order
differential equations ($ B= x + \ii y  $),
\bea {\dot A} & = & 2  {\dot g} = 4 \la f  y \lab{eom1}  \\
{\dot x} & = &  2 y (1 + \la f  ) + 4 \la f x y  \lab{eom2} \\
- {\dot y} & = & 2 x ( 1 + \la f ) + 2 \la f (x^2 - y^2 )  +
\la ( 1 - \e^{2 A} ) /2  \lab{eom3}
\eea
where the dot denotes differentiation with respect to the dimensionless
time scale $\tau = \om t $, one only needs to express $f(t) $ in terms
of $A,x,y$. For this purpose, it is useful to study the averaged
elements of $\CS $ and to commutate $a,\ad $ through $\rho$. Using the
cyclicity of the  trace, this leads to a linear system of equations for
the corresponding averages with the solution
\beq n= {1 \0 Z}   \tr ( \rho \ad a )   =
{ ( 4 r^2 + \e^A - \e^{2 A} )  \0  N }  ,\qd
{1 \0  Z }  \tr ( \rho a^2) = { 2 (x+ \ii y)  \0 N }   \eeq
where $ r^2 =  x^2+y^2 ,\, N= (1 -\e^A)^2 -  4 r^2  ,$
thus $f(t) = (1 -  \e^{2 A} + 4 r^2  + 4 x )/(2 N ) $.

The system (\ref{eom1}-\ref{eom3}) can be solved exactly.
An integral of motion can be obtained by forming $\dot x x + \dot y
y$ and eliminating $f$ by means of  (\ref{eom1}),
\beq r^2 = {1\04} ( 1 + \e^{2 A} + C \e^A  ) , \lab{con1} \eeq
with an integration constant $C$.
A second integral arises from eliminating $y$
\beq \left( \e^{-A} (x+{1\02 } ) + {C\0 4} \right)^2={C+2 \0 2 \la}
\e^{-A} + D \lab{con2} , \eeq
with a second constant of motion  $D$.
Finally, rescaling the thermal average by  $z = - f (C+2 )/2 $, the
time evolution is governed by the first order differential equation
\beq \dot z = \sqrt{ P(z)},\qd P(z)= 1 -{ C^2 \0 4 } + 2 C z -4 z^2
+ {8 \la z \0 C+2}  \left( z^2-D  \right) \label{difeq}   \eeq
which may  be solved in terms of Jacobi integrals.  Note that the
solution of $P(z)=0$  determines the static limit of the average
$\left< X^2 \right>$.

The physical content of (\ref{difeq}) may best be displayed by choosing
particular initial conditions. Consider, e.g., that at $t=0$
the system be prepared in the non-self-interacting ($\la =0$) equilibrium
state, to wit $A(0)=A_0=-\om/T, \, x(0)=y(0)=0,$ corresponding to the
constants of motion
\beq C_0 =-(\e^{A_0} + \e^{-A_0} ),\qd D_0=
{ (1 - \e^{- A_0} )^2 \0 16 \la} \left( 8 + \la ( 1 + \e^{A_0} )^2
\right).
\eeq
In that case
\beq f(t) = { 1  \0 2 ( 1 -\e^{A_0} )^2 } \left( 1-\e^{2 A_0}-u_- \,
\sn^2 ( 2 \Omega  t | m ) \right) \eeq
where $\sn$ is the Jacobian Elliptic Function \cite{abram}
with \beq \Omega=\sqrt{\la u_+ \0 8 } {\om \0 (1- \e^{A_0}  ) }  \eeq
and module
$ m= u_-/u_+ $
where
\bea  u_{\pm} =   { ( 1-\e^{A_0} ) \0 2 \la }   &&
\left(  2  (1-\e^{A_0}  ) + 3 \la (1+ \e^{A_0}  )  \pm \right. \nonumber \\
&& \left. \sqrt{ 4 (1- \e^{ A_0 } )^2 + 12 \la (1-\e^{2 A_0 })
+ \la^2 (1+\e^{A_0} )^2   } \right). \eea
The expectation value $f(t)$
performs an anharmonic oscillation with
period ${\cal P} =   K(m)/ \Omega $
where $K(m)$ is the Complete Elliptic Integral of the first kind
\cit{abram}.
We observe that also the occupation number corresponding to $H_0$
varies in time,
\bea n& =& n_0 + {\la u_- \0 8 ( 1 - \e^{A_0})^4 }\left(2 (1-\e^{2 A_0})
- u_{-} \sn^2 ( 2 \Omega  t | m ) \right) \sn^2 ( 2 \Omega  t | m ) ,
 \nonumber \\
n_0 & = & (  \e^{-A_0} -1 )^{-1}
,\eea
which is not surprising since the particle-number operator $H_0/\om$
no more commutes with the quadratic part of $H$. However,
in the thermal average of $H^{quadr.}$,
\beq h = \li H^{quadr.} \re = \om \left( n(t) + {1 \0 2 } +
{\la \0 2} f(t)^2  \right)  , \eeq
time dependence cancels out in a non-trivial way, as it should be since
we have solved the von~Neuman equation exactly. In fact, the two
integrals of motion (\ref{con1},\ref{con2}) are already sufficient to
determine the energy $h$ of the anharmonic oscillator
\beq h = - {C \0 2 (C+2) } + { 2 \la D \0 (C+2)^2 } \eeq
for arbitrary initial conditions.

It is interesting to study the high temperature limit
where the period approaches
\bea \om {\cal P}  = \sqrt{ \om \pi \0 2\la T }   \pi \lb \Gamma ({3\0 4})
\rb^{-2} -
 \sqrt{ \om \pi \0 2\la T }^{ {}\, 3} \left( \lb\Gamma ({3 \0 4} )\rb^{-2} +
 {1  \0 2 } \lb \Gamma ({ 5\0 4})\rb^{-2} \right) +
{{\cal O}} (T^{-\frac{5}{2}} ), \nonumber \\
 \eea
which is a valid approximation for $2\la T \gg \om \pi $.
The period is a non-analytic function in the
coupling constant in the high temperature limit. Physically this can
be understood as follows. For sufficiently high $T$ the energy
levels with $E_n \sim T$ are all occupied. In that regime the shape
of the potential is dominated by the quartic part, the quadratic
part being rather a perturbation of it.

\section*{Summary and Conclusion}
We studied the dynamics of a thermal non-equilibrium anharmonic
oscillator. Expanding the Hamiltonian around the temperature and time
dependent expectation value $\left< X^2 \right>$, the von~Neuman equation
can be solved exactly for the density operator.  We find that the system
performs an anharmonic oscillation. Specifying the initial conditions to
be the non-interacting thermal equilibrium configuration, the time
evolution is studied in some detail.

The basic ingredient in the calculation is that the
commutator algebra of second order polynomials in
creation/annihilation operators closes. This allows to
commutate operators though the density operator which,
together with the cyclicity of the trace, leads to a set of
linear equations. It would be interesting to extend this method to
go beyond the mean-field approximation which amounts to consider
fourth order polynomials.


\begin{references}
\bibitem{Braa}
E.~Braaten and R.~D.~Pisarski,  Nucl.~Phys.\ {B 337} (1990) 569.
\bibitem{phase}
See e.g.: D.~Boyanovsky, H.~J.~de~Vega and R.~Holman,
{\em Non-equilibrium dynamics of phase transitions: From the early
universe to chiral condensates}, Proceedings of the {\em
2. Journ\'ee Cosmologie}, Observatoire de Paris, June 1994,
H.~J.~de~Vega and N.~S\'anchez, Editors, World Scientific, 1994.
\bibitem{pinch}
A.~J. Niemi and G.~W.~ Semenoff, Ann.~Phys. (N.Y.) 152 (1984) 105;
Nucl.\ Phys.\ B 230[FS] (1984) 181. \\
H.~A.~Weldon, Phys.\ Rev.\ D 45 (1992) 352.\\
T.~Altherr and D.~Seibert, Phys.\  Lett.\ B 333 (1994) 149.\\
P.~F.~Bedaque, Phys.\ Lett.\ B 344 (1995) 23.
\bibitem{coarse}
See e.g.: R.~Zwanzig, in {\em Lectures in Theoretical Physics},
edited by W.~E.~Brittin, B.~W.~Downs and J.~Downs, Interscience, New
York, 1961.
\bibitem{zubarev}
D.~N.~Zubarev, {\em Nonequilibrium statistical thermodynamics},
Consultants Bureau, New York, 1974.
\bibitem{hu}
E.~Calzetta  and B.~L.~Hu, Phys.~Rev.\ D 35 (1987) 495.
\bibitem{balescu}
R.~Balescu, {\em Equilibrium and Nonequilibrium Statistical
Mechanics}, Wiley, New York, 1975.
\bibitem{abram}
M.~Abramowitz and I.~A.~Stegun (eds.), {\em Handbook of Mathematical
Functions}, Dover Publ., New York, 1965.

\end{references}
\end{document}